%% file: paper.tex
\documentclass[letter]{jpsj2} 

\input{header}

\def\vec#1{{\mathbf{#1}}}
\def\paragraph#1{}
\def\i{i}

\title{Nontrivial quantized Berry phases for itinerant spin liquids}
\input{pre-jpsj}

\begin{document}
\maketitle

\paragraph{intro}
New quantum numbers have recently been proposed to classify 
states of matters
beyond the Landau symmetry-breaking description,
such as ground state degeneracy, non-Abelian Berry's phases and edge
excitations\cite{IJMPB.5.1641, PRL.71.3697, JPSJ.73.2604, JPSJ.74.1374}.
Such a nontrivial class of matter 
is identified as possessing topological order,
which 
can be useful to characterize quantum liquids;
the Haldane spin chain\cite{PLA.93.464},
the valence bond solid states\cite{PRL.59.799}, and
the spin-Peierls system\cite{PRL.62.1694}.
Also, the entanglement entropy 
has attracted attention\cite{PRL.90.227902},
which 
provides a new physical insight
for the gapped quantum liquids with a nontrivial
topological order thorough the bulk-edge correspondence
\cite{JPSJ.76.074603, PRB.76.012401}.
Moreover,
topological quantities such as 
the Chern numbers, 
which have successfully characterized
 the quantum Hall states\cite{PRB.40.7387} 
and the generic quantum liquids\cite{JPSJ.73.2604, JPSJ.74.1374},
have the advantage that these quantities are quantized, 
which implies they are topologically protected 
against 
small perturbations.
\paragraph{YH}
In addition to the Chern numbers, the Berry phases can be also quantized in some 
situations.
When the ground state is gapped and is invariant under some anti-unitary operation,
the Berry phase, which takes an arbitrary real number in principle, is quantized to the two values: a trivial value 0 or a nontrivial value $\pi$ (mod $2\pi$).
This scheme has been proposed and applied for
gapped quantum liquids\cite{cond-mat.0603230, JPC.19.145209}.
Since the Berry phase is defined by a {\em local} $SU (2)$ twist at a specific link, 
the Berry phase can be used as a {\it local} order parameter,
which does not have any classical correspondent since it is gauge dependent.
Note that the expectation value of any observable is invariant under unitary transformations of bases, 
which means that it is a gauge invariant.
In dimerized Heisenberg systems,
the Berry phases break the translational symmetry:
the Berry phases at strong bonds (or at weak bonds) are $\pi$ (or zero).
It implies that dimer singlets are mostly localized at the strong
bonds.
The scheme can be applied to the strongly
 correlated electron system as well\cite{cond-mat.0603230}.
\paragraph{Aligia}
To take an advantage of topological stability  against small
perturbations such as randomness, a finite gap above the ground state is required.  
If one does not require the topological stability,
a Berry phase can be obtained even if there is no finite gap\cite{EPL.45.411}.

\paragraph{motivation}
In the present work,
the scheme to calculate the topologically stable Berry phase for a gapped system
is applied to the $t$-$J$ model
as one of the strongly correlated systems\cite{cond-mat.0603230, JPC.19.145209}.
In contrast to the localized singlets in the dimerized Heisenberg systems,
singlets in the $t$-$J$ model are itinerant.
Therefore there can be  gapless charge excitations
 above the ground state,
which implies existence of a low energy cluster of
eigenstates even if the spin excitation is gapped.
Then, we need to use a non-Abelian Berry connection, which is defined 
by all the eigenstates in the cluster\cite{cond-mat.0603230}.
The Berry phase in this case is defined by taking a trace over
the eigenstates.
It
reminds us 
that the standard order parameter as an expectation value of an operator
is calculated by taking a trace with the density matrix when the system is
in a mixed state; besides, when the ground state is degenerate,
one needs to take an average over the degenerate eigenstates at zero temperature.

\paragraph{$t$-$J$ model}
The $t$-$J$ model is an effective model
of the large $U$ repulsive Hubbard model, which describes the hole-doped Mott insulator.
The study on two dimensional(2D) $t$-$J$ models was stimulated 
by the discovery of high-$T_c$ superconductivity of
the copper-oxygen planes\cite{SCIENCE.235.1196,
  PRB.37.3759}.  
In addition, one-dimensional(1D) $t$-$J$ models have received much
attention recently because of  the common properties
between 1D and 2D strongly correlated electron
systems\cite{PRL.64.1839} and, more directly, possibilities of
quasi-1D superconductivity\cite{SCIENCE.271.618}.  Motivated this
proposition, many theoretical studies on 1D $t$-$J$ models were
focused on superconducting states.
For simplicity, 
we focus on the large $J/t$ region in the paper.

\paragraph{Zhang-Rice singlet 'Æ'ÌŠÖŒW}
\paragraph{Hamiltonian}
The $t$-$J$ model is defined on the
subspace without double occupancy as
\begin{eqnarray}
  \label{eq:model}
H &=& H_t + H_J
,
\\
H_t &=& \sum_{ij} t_{ij} {\cal P} \vec{c}^\dagger_{i} \vec{c}_{j} {\cal P}
,
\\
H_J &=& \sum_{ij} J_{ij} \left( \vec{S}_i \cdot \vec{S}_j - {n_i n_j \over 4}\right)
,
\end{eqnarray}
where $\vec{c}_i^\dagger=(c_{i\uparrow}^\dagger, c_{i\downarrow}^\dagger)$, $n_i =  \vec{c}_{i}^\dagger \vec{c}_{i}$, $\vec{S}_i={1\over 2} \vec{c}_{i}^\dagger {\boldsymbol{\sigma}} \vec{c}_{i}$ and ${\boldsymbol{\sigma}}$ is a vector form of the Pauli matrices.
Here the operator ${\cal  P}=\prod_i (1-c_{i\uparrow}^\dagger c_{i\uparrow} c_{i\downarrow}^\dagger c_{i\downarrow})$ is 
the projection to states with no
double-occupancy.
In this paper, we limit ourselves to a 2D square lattice or a 1D chain
under the periodic boundary condition with the lattice size $L$ and the number of up (down) electrons $N_{\uparrow}$($N_{\downarrow}$).
Hereafter, the hopping matrix $t_{ij}$ has a non-zero value $t$ only on nearest-neighbor links,
while dimerization or next-nearest-neighbor interactions can be introduced for the spin exchange $J_{ij}$.
We take $t=1$ as a unit of
energy in the numerical calculations.
\paragraph{Berry phase}
Let us describe the method to calculate Berry phases.
For a parameter-dependent Hamiltonian, $H=H(\theta)$,
the Berry phase $\gamma$ of an $M$-fold multiplet
is customarily given by the $U(1)$ part of non-Abelian Berry connection
as
\begin{math}
\i \gamma = \int_C \mbox{Tr} A(\theta)
\end{math}
,
where the Berry connection $A$ is the $M\times M$ matrix defined as
$\left(A(\theta)\right)_{mm'} = \bra{\psi_m(\theta)}\d{}\ket{\psi_{m'}(\theta)}
=\bra{\psi_m(\theta)}{\d{}\over \d{\theta}}\ket{\psi_{m'}(\theta)} \d{\theta}$
and
states are normalized eigenstates of the Schr\"odinger equation;
\begin{math}
H(\theta) \ket{\psi_m(\theta)} = E_m(\theta) \ket{\psi_m(\theta)}, \;\; (1\leq m \leq M)
\end{math}\cite{PRL.52.2111}.
Note that when the $M$ states are degenerate at $\theta=0$ a standard
order parameter at zero temperature is defined as $\langle {\cal O} \rangle = \sum_m
\bra{\psi_m (0)}{\cal O} \ket{\psi_{m}(0)}/M$ and has some analogy
with the non-Abelian Berry phase.  The main differences are the
differentiation and the integration for the Berry phase, which lead to
the gauge dependence.

The Berry phase $\gamma$ is proved to be real and has ambiguity due to
gauge freedom of eigenstates.
To avoid it,
we should fix the gauge.
Following ref.~\citen{JPSJ.73.2604},
the gauge-fixed states can be obtained from $P\ket{\phi_m}$
with a generic basis set $\ket{\phi_m}$ and the gauge invariant projection operator $P$ defined as $P=\sum_m
\ket{\psi_m (\theta)} \bra{\psi_m (\theta)}$.
However, in this paper, we use another way to calculate the Berry phase
by introducing a {\em gauge-invariant Berry phase } for the lattice analogue
of the Berry connection. 
It is defined as follows
by discretizing the parameter space of 
$\theta$ into $K$ points \cite{cond-mat.0603230, PRB.47.1651}:
\begin{eqnarray}
\gamma  &=&  \lim_{K\rightarrow \infty} \gamma_K,\;\;\;
\gamma_K=-\sum_{k=1}^K \arg \det  C(\theta_k)
\label{eq:berry:lgt}
,
\end{eqnarray}
where $C(\theta_k)$ is the $M\times M$ matrix defined as
\begin{math}
\left( C(\theta_k) \right)_{mm'} = \braket{\psi_m(\theta_k)}{\psi_{m'}(\theta_{k+1})}
\end{math}
under the periodic condition: $\theta_{K+1}=\theta_{1}$.  
The Berry phase in a continuum is gauge dependent but
the one defined here for the  discretized parameter space 
is gauge invariant.
The gauge invariance here means
that it is invariant under gauge transformations after fixing the discretization.
This Berry connection 
can be considered as the connection of 
a 1D analogue of the lattice gauge theory.
Note that the periodic
condition in the parameter space, 
$\ket{\psi_m(\theta_{K+1})}=\ket{\psi_m(\theta_{1})}$,
guarantees the gauge invariance of the Berry phase here.  
\paragraph{Local Spin Twist}
As a generic parameter $\theta$ in the definition of the Berry phase, 
we use a local spin twist in the present study:
\begin{eqnarray}
\label{eq:LocalSpinTwist}
S^+_{i}S^-_{j} + S^-_{i}S^+_{j} &\rightarrow& \e^{\i\theta_{ij}}S^+_{i}S^-_{j}+\e^{-\i\theta_{ij}}  S^-_{i}S^+_{j}
,
\end{eqnarray}
where $S^{\pm}_i = S^x_i \pm \i S^y_i$.
Since the Berry phase is defined by
the local twist at each link $ij$,
the Berry phase is used as a local order parameter on the link.
Note that $\theta_{ij}$
is introduced only to the spin
exchange terms  $H_J$, and it does not modify the other terms in the Hamiltonian.
We denote the Hamiltonian with $\theta_{ij}$ as $H(\theta_{ij})$.

\paragraph{anti unitary operator for $t$-$J$ model}
The gauge-invariant Berry phase here is also
quantized when the Hamiltonian $H(\theta_{ij})$ is
invariant under an anti-unitary operator, i.e., $[H(\theta_{ij}),\Theta]=0$.
The anti-unitary operator in the present case is 
the time-reversal operator 
written as 
$\Theta=KU$
with complex conjugation $K$ and a unitary operator
 defined as $U=\prod_i \e^{{\pi\over 2} (S^+_i - S^-_i)}$.
The quantization of the $K$-discretized Berry phase $\gamma_K$
is proved in the same way as in refs.\citen{cond-mat.0603230} and \citen{JPC.19.145209}.
Since the Berry phase is quantized ($\gamma= 0$ or $\pi$) and is used as a link-variable as discussed above,
each link has one of 
three labels:``$0$-bond'', ``$\pi$-bond'', or ``undefined (or gapless)''.
As a whole system, a texture pattern of the local order parameters is obtained.

\paragraph{numerical error}
Before we show the results,
we comment on some technical aspects.
We can obtain $\gamma$ even for small $K$
without numerical error due to the quantization of $\gamma_K$.
For example, in the one singlet case ($L=2, N_\uparrow=N_\downarrow=1$),
$K=3$ is enough to obtain the correct value, $\gamma=\pi$, in the large $K$ limit.
On the other hand, absolute value of $\displaystyle \Gamma_K=\det \prod_{k=1}^K C(\theta_k)$
is used  as a convenient criterion for the convergence.
We have checked it
for all the results shown below.

\begin{figure}
\resizebox{7cm}{!}{\includegraphics{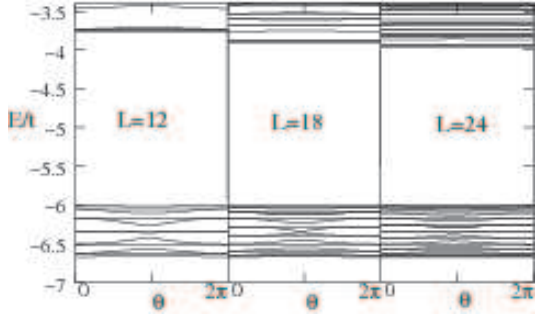}}
\caption{Energy diagram as a function of a local spin twist $\theta$
  in the 1D $t$-$J$ model with $N_\uparrow=N_\downarrow=1$ and $J/t=6$.
The system size is $L=12, 18, 24$.}
\label{fig:A}
\end{figure}

\paragraph{energy digram of perturbation}
In the $t$-$J$ model with a few electrons,
the spin gap is finite
but the charge excitation is gapless.
Then, low energy states below the spin gap
are treated as an $M$-fold multiplet
to calculate the Berry
phase.
To understand the spin gap in a few electrons case,
we consider the large $J$ limit.
For simplicity, let us consider the one singlet case, $N_\uparrow=N_\downarrow=1$
for a while.
The $M$-fold degenerate states below the spin gap form a multiplet,
where $M$ is equal to the number of links
and the spin gap is of the order of $J$.
These are spanned by the localized singlet states
$\ket{l}$,
where the link number $l$ indicates the position of the localized singlet ($1\leq l\leq M$).
These states are bundled as a multiplet.
Then, we consider the hopping process of a singlet as a perturbation in a similar way to refs.~\citen{PRB.37.3759} and \citen{PRL.61.2376}.
In the eigenspace spanned by $\ket{l}$,
the effective Hamiltonian is obtained by the  second perturbation theory as
\begin{eqnarray*}
H_{\rm eff} &=& \sum_{ll'} \ket{l} \bra{l} H^{(2)} \ket{l}\bra{l'}
,\;\;\;
H^{(2)} = H_t {1\over E- H_J} H_t  
.
\end{eqnarray*}
When we switch on hopping process, the
singlet starts to move around 
with the effective hopping $t_s={t^2\over
  J}$.  
The degeneracy is lifted by $H^{(2)}$ and its bandwidth is of the order of $t_s$.  
When $J/t$ is large enough, the spin gap above the band is
stable in the thermodynamic limit, $L\rightarrow \infty$,
as shown in Fig.~\ref{fig:A}.
The charge gap, which is a small gap between the states in the multiplet, is of the order of $t_s/M$
and becomes zero in the thermodynamic limit.
Note that the number of the states is $M=L$ for a 1D chain or $M=2L^2$ for a 2D square lattice.
Although the degeneracy of the multiplet is macroscopic,
the Berry phase of states below the spin gap is  numerically stable.

\begin{figure}
\resizebox{7cm}{!}{\includegraphics{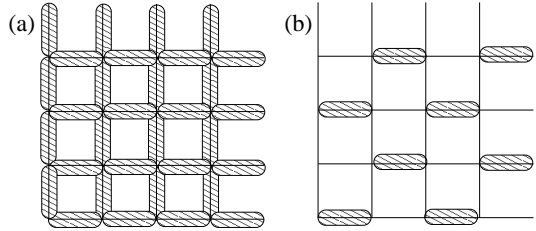}}
\caption{A texture pattern of 
the local order parameters for the 
one singlet states ($N_\uparrow=N_\downarrow=1$) in the 
$t$-$J$ model on a $4\times 4$ lattice under
the periodic boundary condition.
The shadowed links denote the $\pi$-bonds
and the others are the 0-bonds.
(a) The local order parameters for the $M=32$ states for the $t$-$J$ model with  $J/t=8$.
(b) The local order parameters for the $M=8$ states of
the  dimerized $t$-$J$ model with the dimerization $J^S/J^W=2$ and $J^W/t=8$.
The strong bonds with $J^S$ correspond with the $\pi$-bonds and
 the other links are weak bonds with $J^W$.}
\label{fig:B}
\end{figure}

\paragraph{2D}
Figure~\ref{fig:B}(a) 
shows a texture pattern for one singlet states
 ($N_\uparrow=N_\downarrow=1$) in the $t$-$J$ model
 on a $4\times 4$ lattice under the periodic boundary condition.
When the spin exchange $J/t$ is large,
a finite gap exists above the $M=32$ states.
As shown in Fig.~\ref{fig:B}(a) the uniform $\pi$-bonds are obtained.
In addition, when the dimerization of the spin exchange is introduced
as $J^S>J^W$,
 the dimerization gap opens up.
The Berry phase of the states below the dimerization gap
will be $\pi$ for strong bonds with  $J^S$ and 0 for weak bonds with  $J^W$.
 For example, as shown in Fig.~\ref{fig:B}(b),
when eight links with the strong interaction are alternately 
distributed 
along the $x-$axis on a $4\times 4$ lattice,
the dimerization gap opens above the $M=8$ states
and the texture pattern coincides with the distribution of the strong bonds with $J^S$.

\paragraph{a singlet in large $J$ limit}
To explain the  results  in Fig.~\ref{fig:B},
let us consider the perturbation from the large $J/t$ limit in the 
1D case for simplicity.
In the large $J/t$ limit, the Berry phase of the multiplet is given as a sum of respective Berry phases of 
the localized singlet states $\ket{l}$. 
Each state gives the Berry phase $\pi$ only
at the specific link where the singlet exists.
Then the Berry phases of the multiplet are $\pi$ uniformly.
It does not depend on the dimensionality.

\paragraph{to finite large $J$}
Unless the spin gap closes,
 any perturbation cannot modify the texture pattern of the local order parameters 
even if the level-crossing within the multiplet occurs.
This is  the topological stability of the quantized Berry phase.
Then, the result at $J/t=8$ shown in Fig.~\ref{fig:B}(a)
should be the same as uniform $\pi$ Berry phase in the large $J/t$ limit,
because the spin gap does not close at $J/t=8$.
In any dimension, the situation can be the same.
In addition, the result shown in Fig.~\ref{fig:B}(b) is the same as in the case $J^S/t=\infty$ and $J^W/t=0$.

\begin{figure}
\resizebox{7cm}{!}{\includegraphics{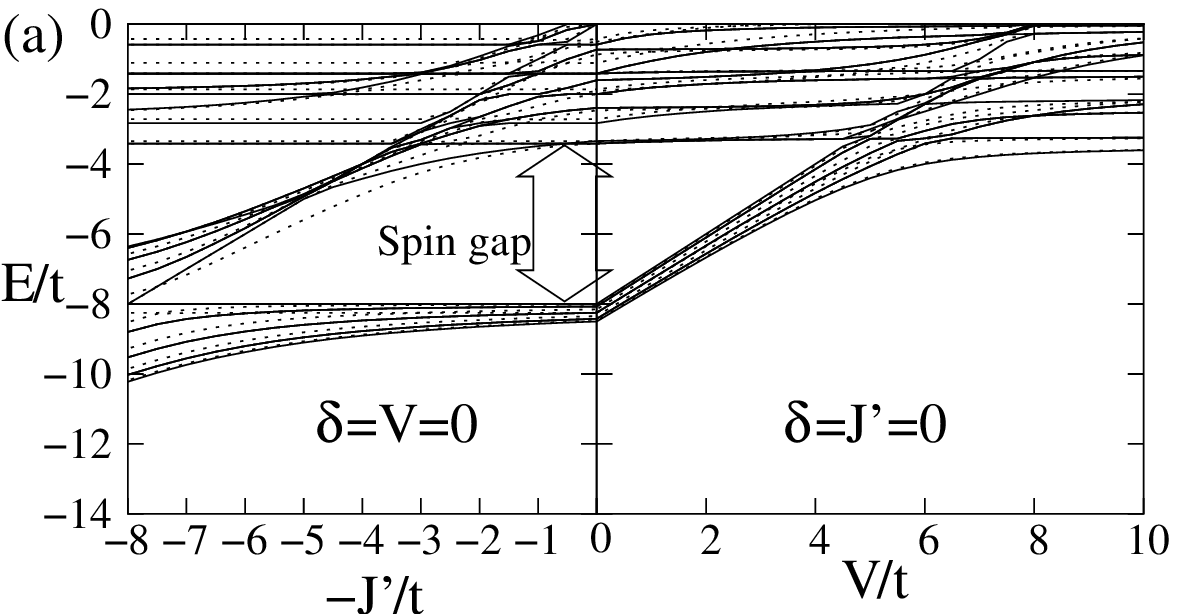}}
\resizebox{7cm}{!}{\includegraphics{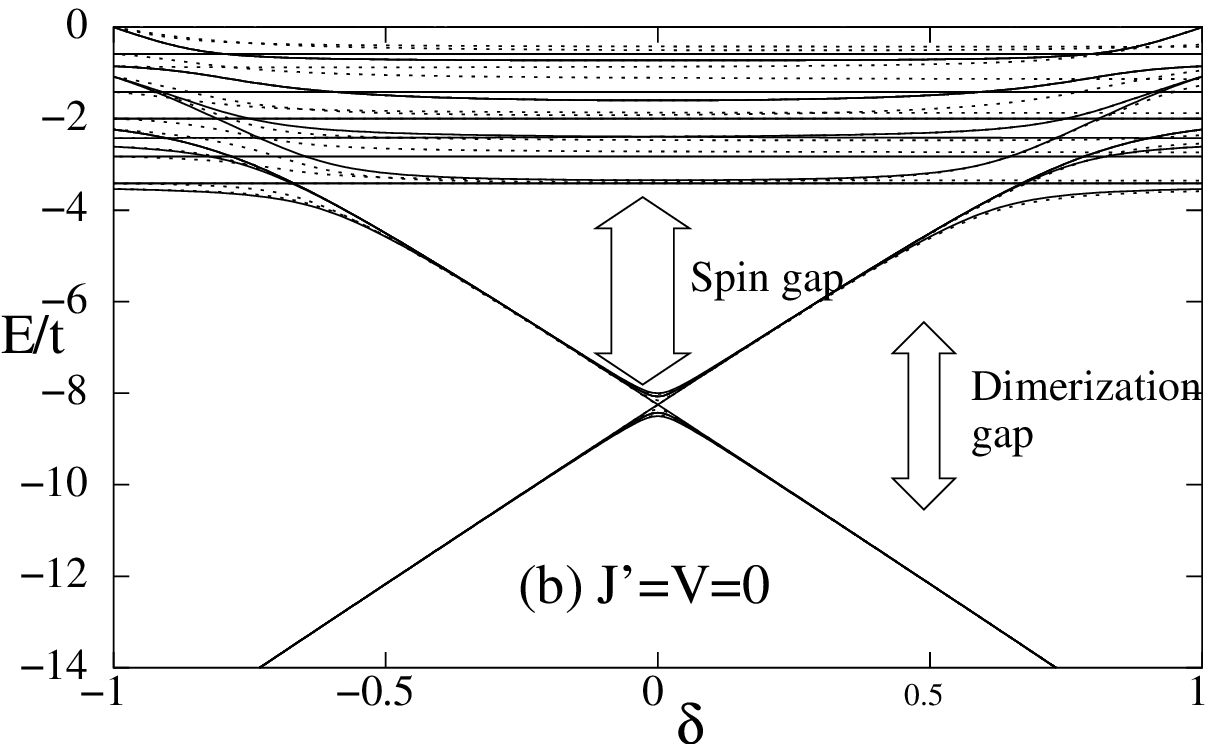}}
\caption{Energy diagram for the 1D $t$-$J$ model with $L=8, J/t=8$
as a function of $J'/t$ or $V/t$ in (a) and $\delta$ in (b).
The solid lines (dotted lines) are energy levels for the local spin twist $\theta=0$ ($\theta=\pi$).
When we consider one of three perturbations of $J'/t$, $V/t$ and $\delta$,
other two parameters are fixed to zero.
The spin gap exists around $E=-8 \sim -4$.
In addition, the dimerization gap opens in (b) at the middle of the multiplet.
}
\label{fig:D}
\end{figure}

\paragraph{another one}
It is interesting to consider the Berry phase in other $t$-$J$ models.
In addition to the simple $t$-$J$ chain,
three modifications can be of interest as spin-gapped systems at finite filling in the 1D case;
(1) the dimerized $t$-$J$ chain\cite{PRB.48.550} by putting $J_{i,i+1} = J (1+
(-1)^i \delta)$, (2) the $t$-$J$-$J'$ chain\cite{PRB.44.12083} by putting
$J_{i,i+1}=J, J_{i,i+2}=J'$ and (3) the $t$-$J$-$V$
chain\cite{PRB.48.4002} by adding $H_V = V\sum_i n_i n_{i+1}$.
We have checked numerically  that small perturbations of
 $\delta$, $J'$, and $V$ do not close
 the spin gap in the $N_\uparrow=N_\downarrow=1$ case
as shown in Fig.~\ref{fig:D}.
The itinerant singlet gives 
uniformly $\pi$ Berry phase as shown 
in the wide class of the $t$-$J$ models.
It should be noted that the dimerization gap opens
 in the middle of the multiplet in the dimerized $t$-$J$ chain
and
the Berry phase of states below the dimerization gap shows the 
simple alternation of $\pi$'s and zeros:
$\gamma=\pi$ (or 0) on a link with the strong (or weak) exchange.
The Berry phase 
gives clear classification of the two topologically different phases
for the dimerized $t$-$J$ chain
as well as for the dimerized Heisenberg chain\cite{JPC.19.145209}.
The differences are (i)the Berry phase in the present paper is that for the  multiplet and 
(ii)the Berry phases of states below the spin gap are $\pi$ uniformly. 

\paragraph{finite filling in 1d}
Let us again consider the $N_\uparrow=N_\downarrow\geq 1$ case in the simple $t$-$J$ chain.
In the large $J/t$ limit, the 
system is in the phase-separated phase and all the electrons  form 
a Heisenberg chain island\cite{PRL.66.2388}.
The length of the island is $N=N_\uparrow+N_\downarrow$
and the translation of this island generates $L$-fold degenerate states.
The Berry phase of each degenerate state
is obtained from that in the Heisenberg chain.
Note that the Berry phases of the  Heisenberg chain 
with even $N$ sites under the open boundary condition 
turn out to be $\pi$ on the left and right boundaries
and show the simple alternation of $\pi$'s and zeros on the other links as mentioned in the dimerized case.
After the summation of respective Berry phases of $L$-fold degenerate states,
it follows that 
the Berry phase of the $L$-fold multiplet in large $J/t$ limit for $N_\uparrow=N_\downarrow=N/2$ and $L>2N$
is uniformly $\pi$ on every links when $N=4n+2$.
It is  uniformly zero when $N=4n$, ($n=0,1,2,\ldots$).
The perturbation with $H_t$ lifts the degeneracy of the large $J/t$ limit.
However, the texture pattern of the Berry phases is
 protected until the spin gap closes.
This is numerically confirmed
for various  $L$ at fixed $N$ in the low density limit.
Note that the spin gap closes at finite filling, i.e. infinite $N$ at fixed $N/L$,
which is the case of usual phase separation in the 1D $t$-$J$ model.

\paragraph{conclusion}
In conclusion, quantized Berry phase as a local order parameter has been
calculated in the 
$t$-$J$ models.
Comparing the previous study in Heisenberg models\cite{cond-mat.0603230, JPC.19.145209},
there are gapless 
charge excitations below the finite spin gap 
which exists in the large $J/t$ case.
Even if the number of states below the gap is large,
the Berry phase has been obtained 
by the exact diagonalization method and Eq.~(\ref{eq:berry:lgt})
successfully.
As a result in the $t$-$J$ models,
the Berry phase for the itinerant singlet
is uniformly $\pi$ on nearest-neighbor links in the 1D and 2D cases.
The itinerant singlet carries the Berry phase $\pi$
in addition to charge, while the singlet does not carry spin.
Due to the topological stability of the quantized Berry phase,
the picture of the 
itinerant singlet obtained in the present study
(for large $J$) has been valid in the wide class of $t$-$J$ models
until the spin gap closes.
The 1D $t$-$J$ model has the uniform Berry phase $\pi$
for the $L$-fold multiplet below the spin gap
especially
when the number of electrons $N $ is $4n +2$.

Moreover, the dimerized $t$-$J$ model has been classified by a texture
pattern of the Berry phases in regard to the dimerization gap.  
In general, the Berry phase can be
defined for each gap and are protected until the gap closing.  
This method will be useful even for the frustrated electron system.
The texture pattern of the Berry phases can be used to find a path to
a simple strong limit without gap closing as the results shown in the
present paper have the corresponding strong limits.  These strong limits
tell us the topological property of the phase by using the adiabatic
continuity.
\acknowledgments 
This work was supported by Grant-in-Aid from
the Ministry of Education, No. 17540347 from JSPS, No.18043007 on
Priority Areas from MEXT and the Sumitomo Foundation.
Some of numerical
calculations were carried out on Altix3700BX2 at YITP in Kyoto
University
and the facilities of the Supercomputer Center, 
Institute for Solid State Physics, University of Tokyo.
\appendix

\input{paper.bbl.save}
\end{document}

%% file: header.tex
\def\d#1{\,{\rm d}#1}
\def\e#1{\,{\rm e}#1}

\def\vec#1{{\overrightarrow{#1}}}

\def\and{\,\,\&\,}

\def\ket#1{|#1\rangle }
\def\bra#1{\langle #1|}
\def\braket#1#2{\langle #1|#2\rangle}

\def\i{{\bf i}}

\def\skipm#1{}


%% file: pre-jpsj.tex
\author{Isao \textsc{Maruyama}$^{1}$\thanks{E-mail address: maru@pothos.t.u-tokyo.ac.jp} and Yasuhiro \textsc{Hatsugai}$^{1,2}$\thanks{E-mail address: hatsugai@sakura.cc.tsukuba.ac.jp}}

\inst{$^{1}$Department of Applied Physics, Univ. of Tokyo, 7-3-1 Hongo, Bunkyo-ku Tokyo 113-8656, JAPAN \\
$^{2}$Institute of Physics, Univ. of Tsukuba, 1-1-1 Tennodai, Tukuba Ibaraki 305-8571, JAPAN
}
\abst{
\input{abst}
}
\kword{Non-Abelian Berry phases, quantum liquids, topological orders, $t$-$J$ model, strongly correlated electron}

%% file: abst.tex
Quantized Berry phases as local order parameters in $t$-$J$ models
are studied.  A texture pattern of the local order parameters is
topologically stable due to the quantization of non-Abelian Berry
phases defined by low-energy states below a spin gap, which exists
in the large $J/t$ case with a few electrons.  We have confirmed
that itinerant singlets in the wide class of $t$-$J$ models carry
the nontrivial Berry phase $\pi$.  In the large $J/t$ case for the
one-dimensional $t$-$J$ model, Berry phases are uniformly $\pi$ when
the number of electrons is $N =4n +2$, ($n=0,1,2,\ldots$).